\documentclass[twocolumn,showpacs,preprintnumbers,amsmath,amssymb,prl]{revtex4}
\usepackage{graphicx}% Include figure files
\usepackage{dcolumn}% Align table columns on decimal point
\usepackage{bm}% bold math

\begin{document}

\preprint{APS/123-QED}

\title{Improved test of Lorentz Invariance in Electrodynamics using Rotating Cryogenic Sapphire Oscillators\\}% Force line breaks with \\

\author{Paul L. Stanwix$^1$}
\email{pstanwix@physics.uwa.edu.au}
\author{Michael E. Tobar$^1$}
\author{Peter Wolf$^{2,3}$}
\author{Clayton R. Locke$^1$}
\author{Eugene N. Ivanov$^1$}
\affiliation{\\ $^1$University of Western Australia, School of
Physics M013, 35 Stirling Hwy., Crawley 6009 WA, Australia\\
$^2$LNE-SYRTE, Observatoire de Paris, 61 Av. de l'Observatoire, 75014 Paris, France\\
$^3$Bureau International des Poids et Mesures, Pavillon de Breteuil,
92312 S\`evres Cedex, France}

\date{\today}% It is always \today, today,

\begin{abstract}
We present new results from our test of Lorentz invariance, which compares two orthogonal cryogenic sapphire microwave oscillators rotating in the lab. We have now acquired over 1 year of data, allowing us to avoid the short data set approximation (less than 1 year) that assumes no cancelation occurs between the $\tilde{\kappa}_{e-}$ and $\tilde{\kappa}_{o+}$ parameters from the photon sector of the standard model extension. Thus, we are able to place independent limits on all eight $\tilde{\kappa}_{e-}$ and $\tilde{\kappa}_{o+}$ parameters. Our results represents up to a factor of 10 improvement over previous non rotating measurements (which independently constrained 7 parameters), and is a slight improvement (except for $\tilde{\kappa}_{e-}^{ZZ}$) over results from previous rotating experiments that assumed the short data set approximation. Also, an analysis in the Robertson-Mansouri-Sexl framework allows us to place a new limit on the isotropy parameter $P_{MM}=\delta-\beta+\frac{1}{2}$ of $9.4(8.1)\times10^{-11}$, an improvement of a factor of 2.
\end{abstract}

\pacs{Valid PACS appear here}% PACS, the Physics and Astronomy
\maketitle

In recent times there has been an increase in activity in experimental tests of Local Lorentz Invariance (LLI), in particular light speed isotropy tests with at least 6 experiments reported in the last 3 years \cite{Lipa,Muller,Wolf04,Antonini,Stanwix,Herrmann}. This is largely due to advances in technology, allowing more precise measurements, and the emergence of the Standard Model Extension (SME) as a framework for the analysis of experiments, providing new interpretations of LLI tests. None of these experiments have yet reported a violation of LLI, though the constraints on a putative violation have become more stringent by approximately three orders of magnitude in the same time frame.

LLI is an underlying principle of relativity, postulating that the outcome of a local experiment is independent of the
velocity and orientation of the apparatus. Tests of LLI are motivated by the central importance of this postulate to modern physics, as well as the development of a number of conflicting unification theories, which
suggest a violation of LLI at some level. To identify a violation it is necessary to have an alternative theory to interpret the experiment \cite{Will}, and many have been developed \cite{Robertson,MaS,LightLee,Ni,Kosto1_1,Kosto1_2,Kosto1_3,KM}. The kinematical frameworks (RMS) \cite{Robertson, MaS} postulate a simple parametrization of the Lorentz transformations with
experiments setting limits on the deviation of those parameters from
their values in special relativity (SR). Due to their simplicity
they have been widely used to interpret
many experiments \cite{Brillet,Muller,WolfGRG,Stanwix,Herrmann,Antonini}. More recently,
a general Lorentz violating extension of the standard model of
particle physics (SME) has been developed \cite{Kosto1_1,Kosto1_2,Kosto1_3} whose
Lagrangian includes all parameterized Lorentz violating terms that
can be formed from known fields. This has inspired a new wave of experiments designed to explore uncharted regions of the SME Lorentz violating parameter space.

Our experiment consists of two cylindrical sapphire resonators of 3 cm diameter and height supported by spindles within super-conducting niobium cavities \cite{Giles}. The sapphire loaded cavities are situated one above the other, oriented with their cylindrical axes orthogonal to each other in the horizontal plane. The experiment is rotated with a period of 18 seconds around its vertical axis.
Whispering gallery modes \cite{wgmode} are excited in each near 10 GHz, with a difference frequency between the two of 226 kHz. The difference frequency along with various experimental parameters are logged by a stationary data acquisition system as a function of the experiments orientation. A detailed description of the experiment can be found in \cite{SR2005}.

Inside the sapphire crystals standing waves are set up with the dominant electric and
magnetic fields in the axial and radial directions respectively,
corresponding to a Poynting vector around the
circumference. The frequency of each resonator $\nu$ is proportional to the speed of light $c$ and inversely proportional to the electrical path length $L$ of the resonator ($\nu \propto c/L$), where $L$ is dependent on the material properties of the sapphire crystal, which have been shown to have a negligible dependence on orientation \cite{MullerPRD}. Hence,  by measuring the difference frequency between the two orthogonal cavities as they rotate we make a direct observation of the isotropy of the speed of light.

To test for Lorentz violations we derive the perturbation of
the difference frequency with respect to an alternative test theory. In the photon sector of the SME
this may be calculated to first order as the
integral over the non-perturbed fields (Eq.(34) of \cite{KM}, see \cite{WolfGRG,SR2005} for an application to our case). The change in orientation of the
fields due to the rotation of the experiment in the lab and Earth's orbital and
sidereal motion induces a time varying modulation of the difference
frequency, which is searched for in the experiment. In the photon sector of the SME \cite{KM}, Lorentz violating terms
are parameterized by 19 independent components, which are in general
grouped into three traceless and symmetric $3\times 3$ matrices
($\tilde{\kappa}_{e+}$, $\tilde{\kappa}_{o-}$, and
$\tilde{\kappa}_{e-}$), one antisymmetric
matrix ($\tilde{\kappa}_{o+}$) and one additional scalar, which all
vanish when LLI is satisfied. The 10 independent components of $\tilde{\kappa}_{e+}$ and
$\tilde{\kappa}_{o-}$ have been constrained by astronomical
measurements to $< 2\times 10^{-32}$ \cite{KM,Kost01}. Recently two combinations of these parameters have been further constrained to less than parts in $10^{-37}$ \cite{Kosto2}. The scalar $\tilde{\kappa}_{tr}$ component has been constrained to $< 10^{-4}$ by \cite{TobarPRD} through the re-analysis of previous Ives-Stilwell experiments, who also propose interferometric techniques to improve on this by seven orders of magnitude. Seven components of $\tilde{\kappa}_{e-}$ and $\tilde{\kappa}_{o+}$ have been independently constrained in stationary optical and microwave cavity experiments \cite{Muller,Wolf04,Lipa} at the $10^{-15}$ and $10^{-11}$ level respectively. The last remaining component $\tilde{\kappa}_{e-}^{ZZ}$ was only recently constrained for the first time by a group of cavity experiments \cite{Stanwix,Herrmann,Antonini,MikeComment,AntoniniComment} designed to both improve on the results of \cite{Muller,Wolf04,Lipa} and, more importantly, be sensitive to $\tilde{\kappa}_{e-}^{ZZ}$ through the use of active rotation in the laboratory.

However, the most stringent independent limits on the isotropy ($\tilde{\kappa}_{e-}$) and boost terms ($\tilde{\kappa}_{o+}$) can only be achieved with 1 year of data. This is because the maximum boost with respect to the Sun Centered Equatorial Celestial Frame (SCECF) is due to the Earth's annual motion. Thus, over 1 year of data is required to decorrelate the parameters. Previous analysis \cite{Lipa,Stanwix,Herrmann}, which contained significantly less than one year of data, constrained the $\tilde{\kappa}_{e-}$ and $\tilde{\kappa}_{o+}$ parameters by assuming no cancelation occurred in the case of a non-zero Lorentz violating effect. We have now acquired sufficient data to remove this assumption, producing independent limits on all of the eight components of $\tilde{\kappa}_{e-}$ and $\tilde{\kappa}_{o+}$.

Alternatively, with respect to the RMS framework, we analyze the change in resonator frequency as a function of the Poynting vector direction with respect to the velocity of the lab in some preferred frame (as in \cite{WolfGRG,SR2005}), typically chosen to be the cosmic microwave background. The RMS parameterizes a possible Lorentz violation by a deviation of the parameters ($\alpha, \beta, \delta$) from their SR values ($-\frac{1}{2}, \frac{1}{2}, 0$). These are typically grouped into three linear combinations representing a measurement of (i) the isotropy of the speed of light ($P_{MM}=\delta -  \beta +
\frac{1}{2}$), a Michelson-Morley (MM) experiment \cite{MM}, constrained by \cite{WolfGRG} to parts in $10^{-9}$ (ii)
the boost dependence of the speed of light ($P_{KT}=\beta - \alpha -
1$), a Kennedy-Thorndike (KT) experiment \cite{KT}, constrained by \cite{WolfGRG} to parts in $10^{-7}$, and (iii) the
time dilation parameter ($P_{IS}=\alpha + \frac{1}{2}$), an
Ives-Stillwell (IS) experiment \cite{IS}, constrained by \cite{Saat} to parts in $10^{-7}$. Because our
experiment compares two cavities it is only sensitive to $P_{MM}$.

In our previous analysis \cite{Stanwix} the amplitude and phase of a Lorentz violating signal was determined by
fitting the parameters of Eq.\ref{nuTest} to the data, with the
phase of the fit adjusted according to the test theory used.

\begin{equation}
\frac{\Delta\nu_0}{\nu_0} = A + B t + \sum_i C_i ~ {\rm
cos}(\omega_{i}t + \varphi_i) +  S_i ~ {\rm sin}(\omega_{i}t +
\varphi_i) \label{nuTest}
\end{equation}

Here $\nu_0$ is the average unperturbed frequency of the two sapphire resonators, and  $\Delta\nu_0$ is the perturbation of the
226 kHz difference frequency. $A$ and $B$ determine the frequency offset
and drift, and $C_i$ and $S_i$ are the amplitudes of a cosine and
sine at frequency $\omega_i$  respectively. In the final analysis we
fit 15 frequencies to the data, $\omega_i = (2\omega_R, 2\omega_R\pm\Omega_\oplus,
2\omega_R\pm\omega_\oplus, 2\omega_R\pm\omega_\oplus\pm\Omega_\oplus, 2\omega_R\pm2\omega_\oplus, 2\omega_R\pm2\omega_\oplus\pm\Omega_\oplus)$, where $\omega_R$ is the rotation frequency of the experiment in the lab and $\omega_\oplus$ and $\Omega_\oplus$ are the sidereal and annual frequencies of the Earth's rotational and orbital motion respectively. Since the residuals of the fit exhibit a significantly non-white behavior,
the optimal regression method is weighted least squares (WLS) \cite{Wolf04}.
WLS involves pre-multiplying both the experimental data and the model matrix by a
whitening matrix determined by the noise type of the residuals of an ordinary least squares analysis. However, this method of analysis proved to be computationally intensive due to the large amount of data we have now acquired. For this reason, an alternative approach used by \cite{Herrmann,Antonini} was adopted. Using this technique we reduce the size of the data set by demodulating it in quadrature with respect to $2\omega_R$ in blocks of 40 periods of rotation. The number of periods was chosen to minimize the net effect of narrow band noise (due to instabilities in the systematic at $2\omega_R$) and broad band noise (due to oscillator frequency noise), which is similar to an optimal filter. By fitting the expression of Eq.\ref{nuTest2} to each block of data using an ordinary least squares regression technique we determine the coefficients $S(t)$ and $C(t)$, which can be considered linear combinations of the sidereal, semi-sidereal, and annual modulations and combinations thereof. The relationship between $S(t)$ and $C(t)$ and the various modulation frequencies is expressed in Eqs.\ref{stage2DAS} and \ref{stage2DAC}, where $\omega_i = (\Omega_\oplus, \omega_\oplus, \omega_\oplus\pm\Omega_\oplus, 2\omega_\oplus, 2\omega_\oplus\pm\Omega_\oplus)$.

\begin{equation}
\frac{\Delta\nu_0}{\nu_0} = A + B t + S(t) ~ {\rm sin}(2\omega_{R} t + \varphi) + C(t) ~ {\rm cos}(2\omega_{R} t + \varphi) \label{nuTest2}
\end{equation}

\begin{equation}
S(t) = S_0 +  \sum_i S_{s,i} ~ {\rm sin}(\omega_i t + \varphi_i)
+  S_{c,i} ~ {\rm cos}(\omega_i t + \varphi_i)
\label{stage2DAS}
\end{equation}

\begin{equation}
C(t) = C_0 +  \sum_i C_{s,i} ~ {\rm sin}(\omega_i t + \varphi_i)
+  C_{c,i} ~ {\rm cos}(\omega_i t + \varphi_i)
\label{stage2DAC}
\end{equation}

A comparison was made between the two techniques by performing a complete analysis of 30 data sets (3 data sets were later excluded from the analysis due to overly large and varying systematic signals at $2\omega_R$). Both techniques produced consistent results, with the uncertainties associated with the demodulated technique being lower than the WLS technique by no more than 15 percent. The difference between the two techniques is most likely due to the efficiency with which the data analysis could be optimized for the noise type present in the data. WLS only takes into account the broad band noise (spectral density) whereas the optimization used in the demodulated technique takes into account the extra noise source of instability of the systematic at $2\omega_R$. Hence, the latter approach was adopted in further investigations of the data.

\begin{figure}
\begin{center}
\includegraphics[width=3.5in]{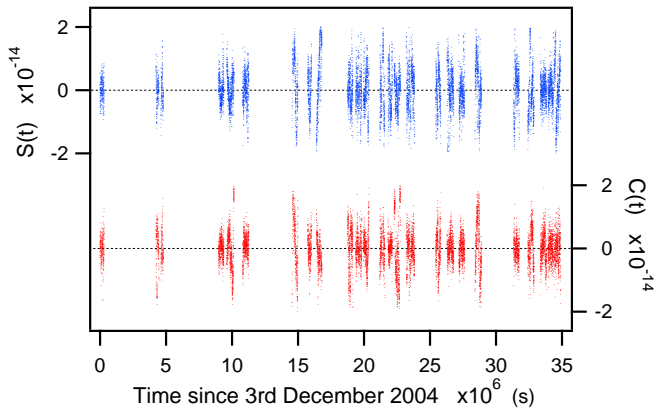}
\caption{Cosine, $C(t)$, and sine, $S(t)$, amplitudes resulting from demodulation of the data at $2\omega_R$ in blocks of 40 rotations, with a linear fit removed from each data set.} \label{fig:data}
\end{center}
\end{figure}

\begin{figure}
\begin{center}
\includegraphics[width=3.0in]{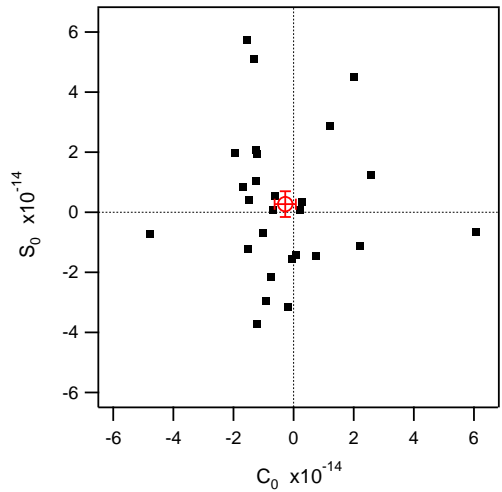}
\caption{The amplitudes $S_0$ and $C_0$ for each of the 27 data sets (squares), used to limit the parameter $\tilde{\kappa}_{e-}^{ZZ}$ of the SME. Also shown (circle) is the mean and its standard error ($S_0=2.9(4.6), C_0=-3.0(3.8)\times10^{-15}$).} \label{fig:SvC}
\end{center}
\end{figure}

The data used in this analysis spans a period from December 2004 to January 2006. It consists of 27 sets of data totalling approximately 121 days. Shown in Fig. \ref{fig:data} are the $S(t)$ and $C(t)$ resulting from the demodulation of the data at $2\omega_R$. An offset and drift has been removed from the coefficients derived from each data set. As described earlier, this data is then used to determine the amplitudes of the frequencies of interest. In \cite{Stanwix} we describe how systematic effects dominate the data at $2\omega_R$, limiting our ability to constrain test theory parameters associated with this frequency (a detailed discussion of the systematics and their effect is thus left out here). Also, we do not consider the nearby annual offsets ($2\omega_R\pm\Omega_\oplus$) for two reasons. Firstly, the strong systematic signal at $2\omega_R$ has been shown to have a significant effect on nearby sidebands due to leakage \cite{Stanwix}, and secondly, by subtracting a linear drift from the individual data sets after being demodulated (as presented in Fig.\ref{fig:data}) it is possible that a signal at the annual frequency may be suppressed. However, all other frequencies of interest (see Tab.\ref{SMETab}) are close to the sidereal or semi-sidereal frequencies, so will be unaffected by the removal of an offset and drift from each data set.

In the SME, all $\tilde{\kappa}_{e-}$ and $\tilde{\kappa}_{o+}$ parameters other than $\tilde{\kappa}_{e-}^{ZZ}$ can be constrained from the sidereal and semi-sidereal frequencies and their annual frequency offsets as outlined in Tab.\ref{SMETab}. $\tilde{\kappa}_{e-}^{ZZ}$ only appears in the coefficient $C_{2\omega_R}$ so to determine a limit we do the same as in \cite{Stanwix} and consider the $C_{2\omega_R}$ coefficients for each data set to be independent and treat them statistically. The systematic at $2\omega_R$ has been shown to be primarily due to tilt variations. It remains relatively constant in phase within a data set but varies between data sets. Fig. \ref{fig:SvC} shows the $C_{2\omega_R}$ and $S_{2\omega_R}$ coefficients for the 27 data sets. Also shown is the mean and standard error of the mean which is used to calculate $\tilde{\kappa}_{e-}^{ZZ}$. The results for the SME analysis are given in Tab.\ref{Results}. We note that the results for $\tilde{\kappa}_{e-}^{XZ}$ and $\tilde{\kappa}_{o+}^{XZ}$ are significant at approximately the $3\sigma$ and $2\sigma$ level  respectively. However, we do not believe this to be an indication of a Lorentz violating effect for reasons similar to those given in \cite{Wolf04}, which also used data taken over more than one year. Our result for $\tilde{\kappa}_{e-}^{XZ}$ is inconsistent with other recent measurements shown in Tab.\ref{Results}. Also, an examination of the corresponding sideband coefficients from an analysis of the individual data sets (not shown here) shows no coherence in the phase of the signal, which would be expected in the presence of a genuine Lorentz violating effect.

\begin{table*}
\caption{\label{SMETab}Shown are the relationships between the $\tilde{\kappa}_{e-}$ and $\tilde{\kappa}_{o+}$ parameters of the SME and the coefficients $C_{C,\omega_i}$,$C_{S,\omega_i}$,$S_{C,\omega_i}$ and $S_{S,\omega_i}$ from Eqs. \ref{stage2DAS} and \ref{stage2DAC} for the 8 frequencies of interest, normalized for the experimental sensitivity $S$. $\chi$ is the colattitude of the lab, $\eta$ is the declination of the Earth's orbit relative to the SCECF, and $\beta_{\oplus}$ and $\beta_L$ are the boost suppression terms due to the lab velocity from Earth's orbital and rotational motion respectively. Also shown is the measured value (in $10^{-16})$ of each coefficient used in the analysis along with its statistical uncertainty. The values for $C_{C,0}$ (used to constrain $\tilde{\kappa}_{e-}^{ZZ}$) and $S_{C,0}$ were determined  by averaging over the data sets (see text). The coefficients of $\Omega_{\oplus}$ were not included in the analysis (see text).}
\begin{ruledtabular}
\begin{tabular}{c|cccc}
  $\omega_i$ & $C_{C,\omega_i}$ & &$C_{S,\omega_i}$ &  \\
\hline

   %Cos
   %***********************************
   %***********************************

   $0$ &
   $\frac{3}{2}{\rm sin}^2({\chi})\tilde{\kappa}_{e-}^{ZZ}$ &
   -30(38) &
   - &
    \\

   $\Omega_{\oplus}$ &
   $-\beta_{\oplus}{\rm sin}^2(\chi)({\rm cos}(\eta)\tilde{\kappa}_{o+}^{XZ}+2{\rm sin}(\eta)\tilde{\kappa}_{o+}^{XY})$ &
 & $-\beta_{\oplus}{\rm sin}^2(\chi)\tilde{\kappa}_{o+}^{XY}$ & \\

   $\omega_{\oplus}-\Omega_{\oplus}$ &
   $\beta_{\oplus}{\rm cos}(\chi){\rm sin}(\chi){\rm sin}(\eta)\tilde{\kappa}_{o+}^{YZ}$ &
   -2.3(0.7)&
   $-2\beta_{\oplus}{\rm cos}(\frac{\eta}{2}){\rm sin}(\chi)({\rm cos}(\frac{\eta}{2})\tilde{\kappa}_{o+}^{XY}-{\rm                     sin}(\frac{\eta}{2})\tilde{\kappa}_{o+}^{XZ})$  &
   0.9(0.7)    \\

   $\omega_{\oplus}$                    &
   ${\rm sin}(2\chi)\tilde{\kappa}_{e-}^{XZ}+2\beta_L\tilde{\kappa}_{o+}^{XZ}$ &
   1.9(0.7)& ${\rm sin}(2\chi)\tilde{\kappa}_{e-}^{YZ}+2\beta_L\tilde{\kappa}_{o+}^{YZ}$ &
    -2.5(0.7)  \\

   $\omega_{\oplus}+\Omega_{\oplus}$    &
   $\beta_{\oplus}{\rm cos}(\chi){\rm sin}(\chi){\rm sin}(\eta)\tilde{\kappa}_{o+}^{YZ}$ &
-2.0(0.7)&
   $-\beta_{\oplus}{\rm sin}(\frac{\eta}{2}){\rm sin}(2\chi)({\rm cos}(\frac{\eta}{2})\tilde{\kappa}_{o+}^{XZ}+{\rm sin}(\frac{\eta}{2})\tilde{\kappa}_{o+}^{XY})$ &
-1(0.7)  \\

   $2\omega_{\oplus}-\Omega_{\oplus}$   &
   $-\frac{1}{2}\beta_{\oplus}{\rm cos}^2\frac{\eta}{2}(3+{\rm cos}(2\chi))\tilde{\kappa}_{o+}^{XZ}$ &
   -0.4(0.7) &
   $-\frac{1}{2}\beta_{\oplus}{\rm cos}^2\frac{\eta}{2}(3+{\rm cos}(2\chi))\tilde{\kappa}_{o+}^{YZ}$ &
   -0.7(0.7) \\

   $2\omega_{\oplus}$                   &
   $-\frac{1}{4}(3+{\rm cos}(2\chi))(\tilde{\kappa}_{e-}^{XX}-\tilde{\kappa}_{e-}^{YY})$ &
   -0.6(0.7) &
   $-\frac{1}{2}(3+{\rm cos}(2\chi))\tilde{\kappa}_{e-}^{XY}$ &
   -1.7(0.7) \\

   $2\omega_{\oplus}+\Omega_{\oplus}$   &
   $\frac{1}{2}\beta_{\oplus}{\rm sin}^2\frac{\eta}{2}(3+{\rm cos}(2\chi))\tilde{\kappa}_{o+}^{XZ}$ &
   -3.4(0.7) &
   $\frac{1}{2}\beta_{\oplus}{\rm sin}^2\frac{\eta}{2}(3+{\rm cos}(2\chi))\tilde{\kappa}_{o+}^{YZ}$ &
   -0.5(0.7) \\

   %Sine
   %***********************************
   %***********************************

   \hline\hline

   &$S_{C,\omega_i}$&&$S_{S,\omega_i}$&\\
   \hline

$0$ &
   $2\beta_L{\rm sin}(\chi)\tilde{\kappa}_{o+}^{XY}$ &
   29(46) &
   - &
    \\

   $\Omega_{\oplus}$ &
   -&
   &
   -&
   \\

   $\omega_{\oplus}-\Omega_{\oplus}$    &
   $\beta_{\oplus}{\rm cos}(\frac{\eta}{2}){\rm sin}(2\chi)({\rm cos}(\frac{\eta}{2})\tilde{\kappa}_{o+}^{XY}-{\rm                     sin}(\frac{\eta}{2})\tilde{\kappa}_{o+}^{XZ})$ &
 -0.3(0.8)&
 $\beta_{\oplus}{\rm sin}(\chi){\rm sin}(\eta)\tilde{\kappa}_{o+}^{YZ}$ &
 -0.8(0.8)  \\

   $\omega_{\oplus}$                    &
    $-2({\rm sin}(\chi)\tilde{\kappa}_{e-}^{YZ}+2\beta_L{\rm cos}(\chi)\tilde{\kappa}_{o+}^{YZ})$&
   1.4(0.8)&
   $2({\rm sin}(\chi)\tilde{\kappa}_{e-}^{XZ}+2\beta_L{\rm cos}(\chi)\tilde{\kappa}_{o+}^{XZ})$ &
   -3.6(0.8) \\

   $\omega_{\oplus}+\Omega_{\oplus}$    &
   $2\beta_{\oplus}{\rm sin}(\frac{\eta}{2}){\rm sin}(\chi)({\rm cos}(\frac{\eta}{2})\tilde{\kappa}_{o+}^{XZ}+{\rm                     sin}(\frac{\eta}{2})\tilde{\kappa}_{o+}^{XY})$ &
 -5.4(0.8)&
 $\beta_{\oplus}{\rm sin}(\chi){\rm sin}(\eta)\tilde{\kappa}_{o+}^{YZ}$ &
 0.4(0.8)  \\

   $2\omega_{\oplus}-\Omega_{\oplus}$   &
   $2\beta_{\oplus}{\rm cos}^2\frac{\eta}{2}{\rm cos}(\chi)\tilde{\kappa}_{o+}^{YZ}$ &
   1.5(0.8) &
   $-2\beta_{\oplus}{\rm cos}^2\frac{\eta}{2}{\rm cos}(\chi)\tilde{\kappa}_{o+}^{XZ}$ &
   -3.2(0.8) \\

   $2\omega_{\oplus}$                   &
   $2{\rm cos}(\chi)\tilde{\kappa}_{e-}^{XY}$ &
   -1.2(0.8) &
   $-{\rm cos}(\chi)(\tilde{\kappa}_{e-}^{XX}-\tilde{\kappa}_{e-}^{YY})$ &
   2.8(0.8) \\

   $2\omega_{\oplus}+\Omega_{\oplus}$   &
   $-2\beta_{\oplus}{\rm sin}^2\frac{\eta}{2}{\rm cos}(\chi)\tilde{\kappa}_{o+}^{YZ}$ &
   0.5(0.8) &
   $2\beta_{\oplus}{\rm sin}^2\frac{\eta}{2}{\rm cos}(\chi)\tilde{\kappa}_{o+}^{XZ}$ &
   3.4(0.8) \\
\end{tabular}
\end{ruledtabular}
\end{table*}

In terms of the RMS framework, the advantage to be gained by having one year of data is primarily statistical. Due to the symmetry of our experiment we are not sensitive to the boost parameter of the RMS, $P_{KT}$, and cavity experiments are not sensitive to the time dilation parameter $\alpha$. However, we can improve on our previous constraint on the isotropy parameter $P_{MM}$ by taking a weighted average over the results of multiple data sets. We analyze each data set using the WLS technique described earlier. The association between $P_{MM}$ and the coefficients of the frequencies of interest is described in \cite{Stanwix}. The coefficients of Eq.\ref{nuTest} are for the frequencies $\omega_i = (2\omega_R,2\omega_R\pm\omega_\oplus, 2\omega_R\pm2\omega_\oplus)$ only. We calculate a value for the RMS parameter of $9.4(8.1)\times10^{-11}$.

In conclusion, by collecting over one year of data we have been able to set the first independent limits on 8 parameters in the photon sector of the SME, without assuming that no cancelation occurs between the isotropy and boost terms. The results do not indicate any Lorentz violating effects, and compared to previous experiments we see a slight improvement in the constraints on these parameters. We improve on our previous determination of $\tilde{\kappa}_{e-}^{ZZ}$ by more than a factor of three. However, due to the systematic disturbances present at twice the rotation frequency we are unable to measure this parameter with the precision of \cite{Herrmann}, who have developed a tilt control system which avoids the major rotation induced systematic. Also, we have reduced the limit on the isotropy parameter $P_{MM}$ of the RMS framework by a factor of two.

To improve on these results we intend to replace the resonators with higher quality sapphire loaded cavities, which have a frequency instability approximately 40 times lower than the current experiment \cite{HartnettEL}. Considerable effort will need to be invested to improve the rotation system and reduce environmental disturbances for this improvement to be realized.

\begin{table}
\caption{\label{Results}Results for the SME Lorentz violation
parameters determined independently in this work. Also shown for comparison is the previous best independent constraints of seven parameters \cite{Wolf04} and more recent short term results that assume no cancelation between the $\tilde{\kappa}_{e-}$ and $\tilde{\kappa}_{o+}$ terms, other than $\tilde{\kappa}_{e-}^{ZZ}$ \cite{Herrmann,Stanwix}($\tilde{\kappa}_{e-}$ in $10^{-16}$, $\tilde{\kappa}_{o+}$ in $10^{-12}$). The $P_{MM}$ parameter from the RMS framework is also listed (in $10^{-11}$). }
\begin{ruledtabular}
\begin{tabular}{c|ccc}
&This work&Previous&Recent short\\
&&analysis \cite{Wolf04}&analysis \cite{Herrmann,Stanwix}\\
\hline
$\tilde{\kappa}_{e-}^{XY}$&2.9(2.3)&-57(23)&-3.1(2.5)\\
$\tilde{\kappa}_{e-}^{XZ}$&-6.9(2.2)&-32(13)&1.9(3.7)\\
$\tilde{\kappa}_{e-}^{YZ}$&2.1(2.1)&-5(13)&-4.5(3.7)\\
$(\tilde{\kappa}_{e-}^{XX}-\tilde{\kappa}_{e-}^{YY})$&-5.0(4.7)&-32(46)&5.4(4.8)\\
$\tilde{\kappa}_{e-}^{ZZ}$&143(179)&-&-19.4(51.8)\\
$\tilde{\kappa}_{o+}^{XY}$&-0.9(2.6)&-18(15)&2.0(2.1)\\
$\tilde{\kappa}_{o+}^{XZ}$&-4.4(2.5)&-14(23)&-3.6(2.7)\\
$\tilde{\kappa}_{o+}^{YZ}$&-3.2(2.3)&27(22)&2.9(2.8)\\
$P_{MM}$&9.4(8.1)&120(220)\cite{WolfGRG}&-21(19)\\
\end{tabular}
\end{ruledtabular}
\end{table}

\begin{acknowledgments}
We would like to thank John Winterflood, Frank van Kann and the technical staff of the School of Physics at UWA for their assistance in this work. This work was funded by the Australian Research Council.
\end{acknowledgments}

\end{document}